\documentclass[12pt,preprint2]{aastex63}
\newcommand{\Ha}{H$\alpha$}

\newcommand{\Hb}{H$\beta$}
\newcommand{\Hgamma}{H$\gamma$}

\newcommand{\heI}{He I}

\newcommand{\kms}{km~s{$^{-1}$}}

\newcommand{\alf}{Alfv\'en}
\newcommand{\cmq}{cm$^{-3}$}

\newcommand{\ozone}{He$\rm ^{+}$+H$^{+}$}

\newcommand{\nzone}{He$\rm ^{o}$+H$^{+}$}

\newcommand{\tC}{{$\theta^1$~Ori~C}}
\newcommand{\tB}{{$\theta^1$~Ori~B}}

\newcommand{\hii}{H II}
\newcommand{\hi}{H I}

\newcommand{\Wfs}{W$\rm _{fs}$}

\newcommand{\Wobs}{W$\rm _{obs}$}
\newcommand{\IF}{H$\rm ^{o}$$\rightarrow$H$\rm ^{+}$}
\newcommand{\oiii}{[O III]}
\newcommand{\oii}{[O II]}
\newcommand{\Cii}{[C II]}
\newcommand{\Siii}{[S III]}
\newcommand{\oi}{[O I]}

\newcommand{\neoniii}{[Ne III]}
\newcommand{\ariii}{[Ar III]}
\newcommand{\nii}{[N II]}
\newcommand{\sii}{[S II]}
\newcommand{\siii}{[S III]}

\newcommand{\feiii}{[Fe III]}
\newcommand{\cliii}{[Cl III]}
\newcommand{\clii}{[Cl II]}

\newcommand{\Ne}{n$\rm_{e}$}

\newcommand{\Vscat}{{\bf V$\rm_{scat}$}}
\newcommand{\Fscat}{{\bf F$\rm_{scat}$}}
\newcommand{\Fmif}{{\bf F$\rm_{mif}$}}

\newcommand{\Vpdr}{{\bf V$\rm _{PDR}$}}

\newcommand{\Vmif}{{\bf V$\rm_{mif}$}}

\newcommand{\Va}{{\bf V$\rm_{A}$}}

\begin{document}

\title{Backscattering and Line Broadening in Orion}

\author{C. R. O'Dell\affil{1}}
\affil{Department of Physics and Astronomy, Vanderbilt University, Nashville, TN 37235-1807}

\author{G. J. Ferland\affil{3}}
\affil{Department of Physics and Astronomy, University of Kentucky, Lexington, KY 40506}

\author{J. E. M\'endez-Delgado\affil{2}\affil{3}}
\affil{Instituto de Astrof\' isica de Canarias (IAC), E-38205 La Laguna, Spain}
\affil{Departamento de Astrof\' isica, Universidad de La Laguna, E-38206 La Laguna, Spain}

\begin{abstract}
Examination of emission lines in high-velocity resolution optical spectra of the Orion Nebula confirms that the velocity component on the red wing of the main ionization front emission line is due to backscattering in the Photon Dominated Region. This scattered light component has a weak wavelength dependence that is consistent with either general interstellar medium particles or  particles in the foreground of the Orion Nebula Cluster. An anomalous line-broadening component that has been known for 60+ years is characterized in unprecedented detail.
Although this extra broadening may be due to turbulence along the line-of-sight of our spectra, we explore the possibility that it is due to \alf\ waves in conditions where the ratio of magnetic and 
thermal energies are about equal and constant throughout the ionized gas. 
%wavelength dependence of backscattering is very different from that found in the reflection nebula surrounding Merope, indicating a difference in the properties of the scattering grains. 
% An anomalous line broadening, that has been known for 60+ years, is most likely due to \alf\ waves. This indicates magneto-hydrodynamics must play a role in photo-evaporation in at least this well studied ionized nebula.
\end{abstract}


\section{Introduction}
\label{sec:intro}

The high surface brightness and proximity of the Orion Nebula (NGC 1976) has attracted hundreds of studies, from x-rays through long wavelength radio rays. These studies have provided a basic model of the bright optical nebula (the Huygens Region), provided tests of basic photoionization theory as it applies to a density range of 10$\rm ^{4}$, and easily observed examples of the interaction of stellar winds and jets with nearby ambient material. Although the associated young stellar cluster is very rich, photoionization is dominated by two early spectral type stars \tC\ and \tB\ whose volumes of dominance are well identified. It is ``everyone's nebula'' and understanding it is a must for understanding other Galactic and Extra-galactic \hii\ regions.

The current study furthers this goal by using the highest velocity resolution optical spectra from published studies and the public domain. These spectra are near the limit of what is possible, as their resolutions resolves the intrinsic line profiles of most of the optical emission lines. We find that there are processes operating (backscattered light and line-broadening) that are usually not considered in the study of other nebulae. We explore these mechanisms in unprecedented detail by employing observations of a wide range of stages of ionization of the emitting gas.

\section{Outline}
\label{sec:outline}

In Section~\ref{sec:background} we summarize earlier work that has lead to this investigation, including discussion of the ionization structure of the Orion Nebula (Section~\ref{sec:zones}), scattered starlight 
(Section~\ref{sec:scatstars}), scattered emission lines (Section~\ref{sec:scatMIF}), and line widths (Section~\ref{sec:broadening}). Section~\ref{sec:existing} presents the data base of spectra used and how they were analyzed. In the discussion section (\ref{sec:discussion}) we analyze the velocities (Section~\ref{sec:velocities}), the properties of the backscattered light forming the red shoulder on the line spectra profiles (Section~\ref{sec:ratio}), and the extra line-broadening that is observed (Sections~\ref{sec:elbc}). Our conclusions are presented in Section~\ref{sec:conclusions}.

\section{Background of this study}
\label{sec:background}

\subsection{Ionization Structure}
\label{sec:zones}

In this study we ascribe the radiation from different observed ions as coming from layered emission zones. This is possible because it is well 
established that the bright Huygens Region is a concave ionized blister \citep{zuk73,bal74,wen95}
  lying beyond the dominant ionizing star \tC\ \citep{ode17b}.

The conditions of ionization are determined primarily by absorption of \tC 's radiation by the strong absorption edges of hydrogen (13.6 eV) and helium (24.6 eV) \citep{agn06}. 
Farthest from \tC\ and the observer will be the hydrogen ionization boundary (IF) where \IF , the limit where the hydrogen is either neutral or ionized. Closer still to \tC\ will be
a boundary where helium begins to be ionized. This means that there will be two ionization zones along a line of sight. The farthest is composed of \nzone\ and the closer is composed of 
\ozone. \tC\ is too cool to produce a He$^{++}$ zone.  Within the allowed zones the stages of ionization of other atoms will be determined by their ionization energies. Oxygen becomes singly ionized by 13.6 eV and doubly ionized by
35.1 eV photons. This means that O$\rm ^{+}$  is found in the \nzone\ zone and O$\rm ^{++}$ in the \ozone\ zone, giving rise to \oii\ in the former and \oiii\ in the latter. Collectively, these zones will be 
called the Main Ionization Front (MIF), thus distinguishing it from scattered light or emission from any foreground layers \citep{ode18}.

This ionization stratification determines the region of emission of the various ions observed in this study. \oi\ arises from exactly at the hydrogen ionization boundary (the {\bf IF}), while  \sii\ arises in a region
near that boundary but within the \nzone\ Zone (a layer we will designated as the ``Near IF Zone or {\bf Near-IF}''. Within the \nzone\ Zone (designated here as the ``Low Ionization Zone'' or {\bf Low-IZ}) will also be found \oii , \nii, \clii, and \feiii\ emission. 
In the \ozone\ zone (designated as the ``High Ionization Zone'' or {\bf High-IZ}) we will find emission from the  \heI\ recombination lines and the \neoniii , \oiii, \siii, and \ariii\ forbidden lines. Of course hydrogen recombination
lines arise both in the \nzone\ and the \ozone\ zones (the ``Hydrogen Emitting Zone'' or {\bf H-Z}) . Dynamic modeling of the nebula \citep{hen03} leads to the expectation that emission closest to the ionization boundary will be close to that of the underlying Photon Dominated Region (27.5$\pm$1.5 \kms ,\footnote{All velocities in this paper are Heliocentric. They may be converted to the LSR system preferred by radio and infrared astronomers by subtracting 18.2 \kms .} \citep{goi15}) and the host Orion Molecular Cloud ({\bf OMC}) 25.9$\pm$1.5 \kms\ \citep{ode18}, while emission from successively higher ionization will be progressively more blueshifted. In the remainder of the paper we give the results for the helium singlet lines ({\bf He-Lines}) separately and also within the {\bf High-IZ}, where the lines originate. This treatment of the hydrogen and helium lines permits the comparison of their recombination line characteristics with the other lines, that arise from collisional excitation.

We give below a listing of the acronyms used in the text. The corresponding regions of emission are ordered with increasing distance along a ray passing from the host molecular cloud towards the dominant ionizing star.

{\bf OMC}: The host Orion Molecular Cloud.

{\bf PDR}: The Photon Dominated Region.

{\bf IF}: The ionization front where \IF .

{\bf Near-IF}: The portion of the \nzone\ Zone near the {\bf IF}.

{\bf Low-IZ}: The \nzone\ Zone.

{\bf High-IZ}: The \ozone\ Zone.

{\bf H-Z}: The hydrogen emitting combined \nzone\ and \ozone\ Zones.

{\bf MIF}: All of the ionized Zones.

{\bf \tC }: The dominant ionizing star.

\subsection{Scattered Starlight}
\label{sec:scatstars}
Beyond the {\bf IF} lies a dust-rich dense layer of atoms and molecules constituting a Photon Dominated Region (PDR) \citep{tie85,tie93}. The dust component can backscatter any optical radiation impinging on it. \citet{green39} first noted that the nebular continuum is much stronger than expected for an ionized gas and attributed this to scattering of \tC\ radiation by dust within the nebula. The excess continuum in several  \hii\ regions, including the Orion Nebula, was reported on in \citet{shajn} where it was concluded that the Orion Nebula's extra continuum was due to scattered starlight. This interpretation was refined as the presence of the PDR was recognized and now we believe that the excess continuum is due to backscattering. More recent studies have quantitatively measured the amount of the excess and the over-all flux distribution of the nebular radiation 
\citep{ode65a,ode10}.

\subsection{Scattered MIF Emission Lines}
\label{sec:scatMIF}
The earliest high spectral resolution studies of the Huygens Region \citep{hoc88,ode92a,jones92} showed a consistent pattern of a red shoulder on emission line profiles. The most probable interpretation was identified in \citet{ode92b}.
Backscattering by grains in the PDR of emission lines from the MIF are to be expected, based on the well established scattering of \tC 's radiation. Because the emitting layers are blue-shifted with respect to the PDR's velocity, the backscattered velocity component will appear as a redshifted shoulder (at velocity \Vscat ) on the MIF's emission at a velocity displacement about twice the velocity difference of the emitting and scattering layers (the photo-evaporation velocity).
This interpretation was confirmed in a modeling study of scattered light by \citet{hen94}, who demonstrated that the double shift was a good upper limit for the scattered light velocity. 

This means that the velocity difference of the MIF component (\Vmif) and that of the PDR (\Vpdr) is a useful measure of the component's photo-evaporation velocity and the difference of velocity \Vscat -\Vmif\  becomes a useful diagnostic of the physical structure of the nebula, especially in the sub-Trapezium region where the MIF					is nearly in the plane of the sky \citep{wen95}. This tool has been used in numerous studies by the lead author of this paper and his collaborators \citep{ode01,abel16,ode17a,ode17b,ode18,abel19,ode20a,ode20b}. 
The ability to accurately measure the backscattered component will primarily depend on the velocity difference (\Vscat - \Vmif), the signal (\Fscat) relative to that of the MIF (\Fmif), that is \Fscat /\Fmif , and the intrinsic line width (usually expressed as the full width at half maximum signal, the FWHM). This means that low mass ions like hydrogen and helium are intrinsically broad through thermal broadening and their \Vscat\ components correspondingly difficult to measure. In contrast, higher mass ions, such those producing \sii\ will be easier to measure. However, in the case of \oi\ and \sii\ the velocity shift \Vscat -\Vmif\ is expected to be small, thus compensating for the narrower line widths, rendering \Vscat\ difficult to measure in these ions. 

The  question of detection is discussed in detail in \citet{ode18}, using a \nii\ line, where detection of a blue-shifted component is tested. Multiple measurements of red components and using the probable errors generated by the profile fitting program (task ``splot'' of IRAF\footnote{IRAF is distributed by the National Optical Astronomy Observatories, which is operated by the Association of Universities for Research in Astronomy, Inc.\ under cooperative agreement with the National Science foundation.}) shows that the threshold of detectability is \Fscat /\Fmif $\approxeq$0.03. Ratios below this level were not used in the current study. This turned out to only apply to \Fscat /\Fmif\ in the \oi\ and \sii\ lines.

Our study determines the \Fscat /\Fmif\ ratio for an unprecedented number of emission lines, covering a unique number of ionization states. One of our goals has been to use this ratio to study the structure of the MIF emission and to possibly determine the wavelength dependence of the backscattering.

\subsection{Line Broadening}
\label{sec:broadening}

At the same time that the early high  spectral resolution spectra revealed the presence of a red shoulder due to back-scattering, it was also noted that the line widths were of unexpected width.
This was first noted earlier in a high-resolution optical photographic study by \citet{ocw59}, where the extra width was assigned to turbulence \citep{mun58}. If this component is due to mass motion, %for the heavier ions 
it carries as much energy as the thermal motion of the gas. Although known for over one-half century, this component has defied a clear interpretation. The most complete discussion of the Extra Line Broadening Component (ELBC) and its
interpretation appears in \citet{ode17b}. Our study extends that investigation, using a wider variety of emission lines.

The observed FWHM (\Wobs) will be the quadratic addition of the thermal broadening, the resolution of the spectrograph being used (R), and any fine structure component (\Wfs).  
Anticipating that this study will reveal an additional line-broadening component, called in \citet{ode17b} the ELBC, this component must be 
included. Using the characteristic electron temperature of 9200 K \citep{md21},  the expected observed  \Wobs\ in \kms\ will be

%\Wobs\ = [424.1/A +ELBC$\rm^{2}$ + R$\rm ^{2}$ +\Wfs$\rm ^{2}$]$\rm ^ {1/2}$
%\begin{equation}   Format from Eduardo. It prints out the same as mine, as written below.
%\label{eq:Wobs}
%\text{W}_{\text{obs}}\ = \left[424.1/\text{A} +\text{ELBC}^{2} + \text{R}^{2} +\text{W}_{\text{fs}} ^{2} \right]^ \frac{1}{2}
%\end{equation}
\begin{equation}
\label{eq:Wobs}
\rm W_{obs} = \left[438.1/A +ELBC^{2} + R^{2} +W_{fs}^{2} \right]^ \frac{1}{2}
\end{equation}
where A is the atomic mass of the emitting ion. The 438.1/A term arises from thermal broadening and is proportional to the electron temperature.
If one ignores the ELBC component, the temperature dependence of the relation allows one to 
derive an electron temperature by observing the FWHM of two lines of very different masses. This has typically 
been doing using \Ha\ or \Hb\ lines together with the strong \oiii\  or \nii\ lines. Because the ELBC component can be important, the derived temperatures are incorrect. \citet{md21} tried to improve this method of temperature determination using assumptions about the nature of the ELBC, following a procedure introduced in \citet{gar08}. 

Even in those lines where \Fscat /\Fmif\ cannot be determined, \Wobs\ can be accurately derived, thus allowing the determination of the ELBC. One of the main contributions of the present study is the 
measurement of many lines originating from multiple ionization zones, which allows an unprecedented study of the line-broadening.  

\section{Analysis of Existing Spectra}
\label{sec:existing}

We have drawn on two sets of existing slit spectra. 

\begin{figure}
\includegraphics
[width=\columnwidth]
{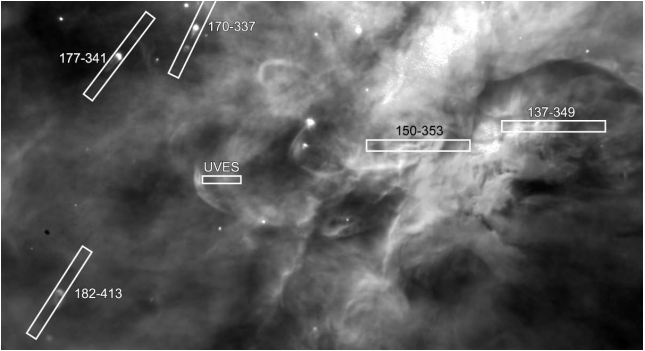}
\caption{This HST F656N WFPC image is 86\farcs8 $\times$47\farcs2 and centered on 5:35:15.82 -5:23:57.0 (2000). It shows the location of the samples used in this study (Section~\ref{sec:existing}). }
\label{fig:locations}
\end{figure}

The first source is from the study of \citet{md21}, which was an echelle slit investigation of the high-ionization shocks that compose HH 529. Since we were interested in the nebula's radiation, we selected a 5\farcs2 long and 1\farcs0 wide sample at 5:35:16.73 -5:23:57\farcs6 free (except at large blueshifts) of shock emission.  This spectrum was made with the UVES spectrograph of the Very Large Telescope (VLT).
    The position of our sample is shown in Figure~\ref{fig:locations}, where it is labelled UVES. The HH~529 shocks lie well removed in velocity from the layers producing the ambient emission-line spectra of this region and are created in the {\bf High-IZ} cavity near \tC\ \citep{bla07}. Aside from these foreground shocks, the MIF in this region is quiescent, as shown in Figure~15 of \citet{gar08}.

\begin{figure}
 \includegraphics
[width=\columnwidth]
%[width=7.5in]	
 {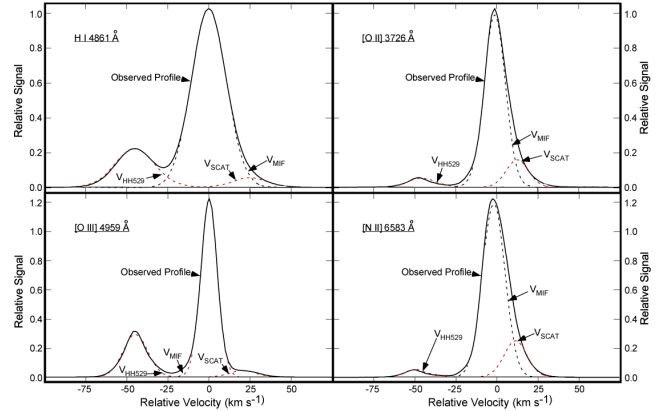}
\caption{This figure shows the observed profiles for four UVES lines. The upper panel is for the \Hb\ line and the lower is for the \oiii 4959 \AA\ line. The velocity scale is relative to \Vmif\ for that line as given in Table~\ref{tab:AllUVESLines}. Solid lines show the observed line profiles while the dashed red lines give the three component fits obtained with IRAF task ``splot''. The blue-shifted component labeled HH~529 is from one of that object's shocks. They are relatively weaker for lower ionization MIF lines. They are not important in analysis of the \Vscat\ components because their back scattered components would be at about 70 \kms.}
\label{fig:profiles}
\end{figure}

The second source was Keck-I Telescope HIRES spectra made with 14\arcsec\ slits centered on proplyds and other regions of particular interest. Portions of the slits were sampled avoiding the proplyds or when the large blue-shift emission could be avoided. They are described in \citet{hen99}. Each of these subsamples were observed several times, so that there are typically about twenty spectra for each pointing. 
The full length entrance slits are shown in Figure~\ref{fig:locations} and are labeled with the targeted object's position labeled with the position in the system introduced in \citet{ode92a}. 

The spectra were resolved into multiple velocity components using the IRAF package of data processing tools, especially the task ``splot'' as was done in multiple previous studies.Often only an assumption of two 
components (MIF the strongest and the much weaker backscattered component). In some cases a well-separated blue component arising from the jets and shocks was also fit. Task ``splot'' yields the signal, velocity, and FWHM for each component and its probable errors. An illustration of an ``splot'' analysis of a narrow and a broad emission line is shown in Figure~\ref{fig:profiles}. 
Comparison of the \hi\ 4861 \AA\ and \oiii\ 4959 \AA\ lines shows the advantages in measuring the \Vscat\ component in lines with a small thermal broadening component. Even though the velocity separation of \Vmif\ and \Vscat\  is smaller for the \oii\ 3726 \AA\ doublet lines and the \nii\ 6583 \AA\ line, the right hand panels show that their \Fscat /\Fmif\ ratios are large and measurable. This significance of this result is discussed in detail in Section~\ref{sec:ratio}.

\begin{deluxetable}{lccccc}
\setlength{\tabcolsep}{0.02in}
\tabletypesize{\scriptsize}
\tablecaption{Data for all UVES Sample Lines*$\dagger$
\label{tab:AllUVESLines}}
\tablewidth{0pt}
%\tablewidth{column}
\tablehead{
\colhead{Ion}&
\colhead{$\rm \lambda _{o}$ (\AA)} &
\colhead{\Vmif } &
\colhead{\Vscat - \Vmif} &
\colhead{\Fscat / \Fmif} &
\colhead{ELBC}}
\startdata
{\bf IF}      &---          &---      &---     &---    &---  \\
\oi               &6300.304 &28.1$\pm$0.0  &---                    &---                        &7.9$\pm$0.2   \\
\oi               &6363.776 & 28.1$\pm$0.1 &---                    &---                        &8.1$\pm$0.0  \\
{\bf Near-IF} &---          &---      &---     &---    &---  \\
\sii               &4068.600 &22.3$\pm$0.0  &---                    &---                      &16.2$\pm$0.2  \\
\sii               &6716.440 &23.4$\pm$0.0  &---                    &---                      &19.5$\pm$0.1 \\
\sii              &6730.820  &23.4$\pm$0.0  &---                     &---                     &18.5$\pm$0.1 \\
\sii             &10320.490 &21.2$\pm$0.0  &---                     &---                      &19.6$\pm$0.5  \\
\sii             &10336.410 &21.4$\pm$0.0  &---                     &---                      &19.9$\pm$0.6 \\
\clii            &9123.600    &24.1$\pm$0.2  &---                     &---                       &18.7$\pm$0.7 \\
{\bf Low-IZ} &  ~       &     ~                  &  ~                    &      ~                  &      ~              \\          
%~~~~~Zone&---&---                    &---                     &---                      &---                   \\
\oii              &3726.030    &18.3$\pm$0.0  &13.4$\pm$0.0  &0.18$\pm$0.01  &7.4$\pm$0.1\\
\oii              &3728.820    &15.1$\pm$0.0  &14.1$\pm$0.0  &0.17$\pm$0.00 &12.6$\pm$0.0\\
\nii              &6548.050    &20.7$\pm$0.0  &13.4$\pm$0.0  &0.17$\pm$0.00 &13.6$\pm$0.1   \\
\nii              &6583.450    &18.4$\pm$0.0  &12.9$\pm$0.0  &0.21$\pm$0.00 &13.2$\pm$0.1   \\
%\Heo$\rightarrow$\Heplus\ Boundary   &---     &---  &---     &---     &--- &--- &---     &---\\
\feiii **           &4658.10      &17.9$\pm$0.4  &12.8$\pm$0.1  &0.19$\pm$0.00  &8.6$\pm$0.1      \\
\feiii **          &4881.11       &12.7$\pm$0.5  &12.3$\pm$0.5  &0.20$\pm$0.01  &8.9$\pm$0.3       \\
%\cliii             &5517.709     &14.3$\pm$0.  &16.0    &0.11   &10.2\\
%\cliii             &5537.873     &14.3  &18.3    &0.09   &10.2\\
{\bf High-IZ} &---             &---                    &---                   &---                        &---                        \\
%~~~~~Zone   &---             &---                    &---                   &---                        &---                        \\
\neoniii            &3868.75 &15.8$\pm$0.0 &17.0$\pm$0.0  &0.08$\pm$0.00   &7.2$\pm$0.1  \\
\neoniii            &3967.46 &15.5$\pm$0.0 &22.1$\pm$0.3 &0.06$\pm$0.00   &7.9$\pm$0.1  \\
\oiii              &4363.21     &14.7$\pm$0.0 &23.1$\pm$0.9 &0.06$\pm$0.00   &9.1$\pm$0.1  \\
\oiii              &4958.910   &16.4$\pm$0.0 &21.6$\pm$0.1 &0.05$\pm$0.00   &7.2$\pm$0.1  \\
\oiii              &5006.840   &17.4$\pm$0.0 &23.1$\pm$0.1 &0.06$\pm$0.00 &6.7$\pm$0.1  \\
\siii              &6312.070   &16.3$\pm$0.0 &17.9$\pm$0.1 &0.11$\pm$0.00    &10.3$\pm$0.2 \\
%\siii                  &9068.930     &17.0 &19.7    &0.10   &9.7\\
\siii                  &9530.980     &15.8$\pm$0.0 &21.2$\pm$0.11&0.10$\pm$0.00 &9.9$\pm$0.1  \\
\ariii                &7135.790      &15.2$\pm$0.0 &16.7$\pm$0.0&0.09$\pm$0.00 &6.3$\pm$0.1  \\
\ariii                 &7751.100     &16.1$\pm$0.0 &18.6$\pm$0.1&0.08$\pm$0.01&6.5$\pm$0.2  \\
{\bf He-Lines}         &---         &--- &--- &---       &---                 \\
He                  &3964.730    &15.8$\pm$0.0 &22.1$\pm$1.6  &0.05$\pm$0.01  &9.5$\pm$0.2  \\
He                  &4921.930    &15.8$\pm$0.0 &19.6$\pm$0.5&0.10$\pm$0.00 &7.8$\pm$0.1  \\
He                  &5015.680    &16.3$\pm$0.0 &19.0$\pm$0.5  &0.10$\pm$0.00 &6.7$\pm$0.2  \\
He                  &6678.150    &16.0$\pm$0.0 &20.0$\pm$0.2  &0.06$\pm$0.00 &7.5$\pm$0.3  \\
He                  &7281.350    &16.5$\pm$0.0 &19.4$\pm$0.0&0.10$\pm$0.00 &7.3$\pm$0.0 \\
{\bf H-Z}      &---         &--- &--- &---      &---                  \\
H9                   &3835.4        &14.3$\pm$0.0 &25.9$\pm$0.8  &0.05$\pm$0.00 &11.6$\pm$0.2  \\
H7                  &3970.08      &15.4$\pm$0.0 &26.8$\pm$0.2  &0.04$\pm$0.00&11.0$\pm$0.1   \\
H6                   &4101.71     &17.4$\pm$0.0 &27.5$\pm$0.1 &0.08$\pm$0.00 &11.5$\pm$0.1  \\
\Hgamma        &4340.47      &15.0$\pm$0.0 &24.2$\pm$0.1  &0.05$\pm$0.00 &10.2$\pm$0.1*** \\
\Hb                  &4861.35     &15.8$\pm$0.0 &23.6$\pm$0.1 &0.06$\pm$0.00 &9.8$\pm$0.1*** \\
\Ha                  &6562.79      &16.0$\pm$0.0 &25.2$\pm$0.1   &0.05$\pm$0.00 &11.0$\pm$0.1***\\
Pa12               &8750.46      &16.7$\pm$0.0 &23.6$\pm$0.0  &0.07$\pm$0.00 &9.4$\pm$0.1  \\
Pa11               &8862.89      &12.6$\pm$0.0 &23.0$\pm$0.1&0.09$\pm$0.00&9.2$\pm$0.0  \\
\enddata

~*~All velocities are in \kms\ and Heliocentric for \Vmif.

$\dagger$~rms indicates the root-mean-squared deviation of a value, as determined from the IRAF task ``splot''.

**~Not included in the average for the {\bf Low-IZ} \Vmif\ values, as explained in Section~\ref{sec:velocities}.

***~ Includes correction of line width for fine structure broadening, using values from \citet{gar08}.

\end{deluxetable}

For the UVES samples the derived values appear in Table~\ref{tab:AllUVESLines}. In that table ions appear grouped into the ionization zones described in Section~\ref{sec:zones} and are ordered with increasing wavelength for each ion. Each column ordered left to right presents: one (ionization zone and ion), two (the assumed at rest wavelength of the emission line), three (\Vmif), four (\Vscat -\Vmif), five (\Fscat /\Fmif), six (the derived ELBC, using observed the FWHM, an assumed electron temperature of 9200 K \citep{md21}, fine-structure broadening components from \citet{gar08}, and a resolution of R=7.5 \kms. The probable errors are those given by ``splot''. It should be noted that no \heI\ triplet lines appear because they have strong fine-structure broadening, something absent in the singlets.

The same results are given given in Table~\ref{tab:AllHIRESLines} for the HIRES observations. The results are ordered in terms of decreasing \Vmif .
There were about 20 samples measured for each line and we show uncertainties determined by the spread of values within each line. 
 Derivation of the ELBC component was the same as for the UVES spectra, but now a resolution of R = 6.2 \kms\ was adopted. 
 
In Table~\ref{tab:AllUVESLines} we have placed the results for \clii\ in the {\bf Near-IF} because the transition from Cl$\rm ^{o}$ to
Cl$\rm ^{+}$ occurs at 13.0 eV and from Cl$\rm ^{+}$  to Cl$\rm ^{++}$ at 23.8 eV. This means that some or most of the \clii\ emission occurs at or near the IF, where the velocity will be that of the {\bf IF} or {\bf Near-IF}. Like the \sii\ emission lines, no \Vscat\ component was detected.

\begin{deluxetable}{lcccc}
\setlength{\tabcolsep}{0.02in}
\tabletypesize{\scriptsize}
\tablecaption{Averaged$\dagger$  Data from Keck-I HIRES Observations*
\label{tab:AllHIRESLines}}
\tablewidth{0pt}
\tablehead{
\colhead{Line}&
\colhead{\Vmif} &
\colhead{\Vscat - \Vmif} &
\colhead{\Fscat/\Fmif}    &
\colhead{ELBC}}
\startdata
\oi\ 6300 \AA               &27.7$\pm$1.8**   &---                    &---                     & 10.3$\pm$1.8\\
\sii\ 6731 \AA              &23.7$\pm$1.9   &15.5$\pm$3.2 &0.08$\pm$0.04 &13.6$\pm$2.5\\
\nii\ 6583 \AA              &20.9$\pm$1.8  &14.4$\pm$1.8 &0.09$\pm$0.04 &11.1$\pm$1.6\\
\oiii\ 5007 \AA             &20.7$\pm$1.9  &14.0$\pm$4.5 &0.06$\pm$0.04  &11.2$\pm$2.7\\
\oiii\ 4959 \AA             &17.7$\pm$2.0 &15.8$\pm$4.3  &0.06$\pm$0.03  &10.8$\pm$2.5\\
\Ha\ 6563 \AA             &17.2$\pm$3.5 &23.2$\pm$3.9  &0.11$\pm$0.06  &---$\dagger$$\dagger$\\ %19.7$\pm$3.4$\dagger$$\dagger$\\
\Hb\ 4861 \AA             &15.3$\pm$3.5 &24.1$\pm$3.8  &0.09$\pm$0.07  &---$\dagger$$\dagger$\\ %18.3$\pm$4.1$\dagger$$\dagger$ \\
\enddata
%round off ratios to hundredths, do better ids on H
~*~All velocities are in \kms\ and Heliocentric for \Vmif.

$\dagger$ ~Averaged over five slits and about 20 samples for each line.

$\dagger$$\dagger$ ELGC not calculated because most of the line width is due to thermal broadening.

**All rms errors are determined from the scatter of about twenty samples for each line.
\end{deluxetable}

\section{Discussion}
\label{sec:discussion}
In this section we discuss what can be learned from the \Vmif\ values, the \Fscat /\Fmif\ ratio, and the ELBC values.
\subsection{Velocities}
\label{sec:velocities}

\begin{deluxetable}{lcccc}
%\floattable
\setlength{\tabcolsep}{0.02in}
\tabletypesize{\scriptsize}
%[width=4 in]
\tablecaption{Averaged Data for UVES Zones*
\label{tab:AllUVESzones}}
\tablewidth{0pt}
\tablehead{
\colhead{Group}&
%\colhead{$\rm \lambda _{o}$ (\AA)} &
\colhead{\Vmif} &
\colhead{\Vscat - \Vmif} &
\colhead{\Fscat/\Fmif}    &
\colhead{ELBC}}
\startdata
{\bf IF}  (\oi)                                &28.1$\pm$0.1   &---                    &---                      &8.0$\pm$0.1\\
{\bf Near-IF} (\sii +\clii)                  &22.6$\pm$1.2   &---                   &---                       &18.7$\pm$1.3\\
{\bf Low-IZ}                        &18.1$\pm$2.3  &13.2$\pm$0.6 &0.19$\pm$0.02 &11.9$\pm$3.9\\
%\Heo $\rightarrow$\Heplus\ Boundary&18.2$\pm$3.7  &$\pm$2.1  &17.5$\pm$1.0 &0.10$\pm$0.01  &10.4$\pm$0.2\\
{\bf High-IZ}                         &15.9$\pm$0.8 &20.1$\pm$2.6  &0.08$\pm$0.02  &7.9$\pm$1.5\\
{\bf He-Lines}                          &16.1$\pm$0.3 &20.0$\pm$1.2  &0.08$\pm$0.02  &8.2$\pm$1.9\\
{\bf High-IZ}+{\bf He-Lines}   &16.0$\pm$0.6 &20.1$\pm$2.1  &0.08$\pm$0.02  &8.0$\pm$1.6 \\
{\bf H-Z}                        &15.4$\pm$1.5 &25.0$\pm$1.6  &0.06$\pm$0.02  &10.5$\pm$0.9\\
 \enddata
%round off ratios to hundredths, do better ids on H
~*All velocities are in \kms\ and Heliocentric for \Vmif.\\
\end{deluxetable}

Because the velocity of the PDR must vary across the Huygens Region, as the local tilt of the PDR changes \citep{ode18}, we limit our interpretation of the velocities to the single sample in the UVES observations.
These are summarized in Table~\ref{tab:AllUVESzones}. For reference, the average \Vpdr\ for the Huygens Region as determined from \Cii\ 158 $\mu$m is 27.5$\pm$1.5 \kms\ \citep{goi15} and is indistinguishably  the same in the region of the UVES observations.  CO emission must occur slightly further into the host cloud and in table 3.3.VII of \citet{gou82} the average velocity is 27.3$\pm$0.3 \kms. We have grouped the results by their ionization zones, as explained in Section~\ref{sec:zones}. This is more meaningful than presenting them by the ion's ionization energies, as frequently done. Ionization energies are what determine the region of emission, which in turn then determines the velocity, rather than the ionization energies directly determining the velocity.

Because the ionized gas nearest to the {\bf IF} is higher density than gas further away, first order considerations lead to the expectation that the higher ionization lines will have more negative velocities. This was the assumption that led \citet{zuk73} and \citet{bal74} to establish that the Huygens Region was a more distant thin emitting layer, rather than a basically symmetric distribution of clumped gas around \tC. This expectation was confirmed in detailed modeling by \citet{hen05}. 

This sense of variation in \Vmif\ agrees with the values in Table~\ref{tab:AllUVESzones}.
% with the exception of the {\bf Low-IZ} where the average (18.1$\pm$2.3 \kms) has a large uncertainty because of the spread in \Vmif\ values. The \clii\ value of 24.1$\pm$0.2 \kms\ is unexpectedly high.  
The two \feiii\ \Vmif\ values are very different, 17.9$\pm$0.4 \kms\ for the stronger 4658 \AA\ line and 12.7$\pm$0.5 \kms\ for the 4881 \AA\ line. These lines both result from collisional excitations out of the 3d$\rm ^{6}$\ $\rm ^{5}$D$\rm_{4}$ ground state of Fe$\rm ^{++}$. The difference is probably due to uncertainties in their intrinsic wavelengths. This is recognized in \citet{md21} and the upper right panel of their Figure~13 shows that the \Vmif\ for this ion is about 17 \kms, consistent with the {\bf Low-IZ} average of 18.1$\pm$2.3 \kms . The uncertainty in intrinsic wavelength of each line does not affect their  \Vscat -\Vmif\ values and we have used them in deriving that average .

In Table~\ref{tab:AllUVESzones} we see a steady blue-shift progression of \Vmif\ values from {\bf IF} through the {\bf High-IZ}, as predicted by the \citet{hen05} study. 
The {\bf He-Lines} value of 16.1$\pm$0.3 \kms\ lies within the value of 15.9$\pm$0.8 \kms\ for the {\bf High-IZ}, as it should be according to the arguments presented in Section~\ref{sec:zones}. The only expected difference would be due to the fact that the collisionally excited forbidden lines used in calculating the average for {\bf High-IZ} would selectively come from high-temperate gas along the line-of-sight, while the recombination \heI\ lines would come from any lower temperature gas. 

Hydrogen recombination lines will arise from all of the ionization zones, their total emission being redder close to the high density {\bf IF} and bluer from any low electron temperature components. At \Vmif\ = 15.4$\pm$1.5 \kms\ its value is indistinguishable from the {\bf High-IZ}+{\bf He-Lines} value of 16.0$\pm$0.6 \kms 

As noted in Section~\ref{sec:scatMIF} the expectation for \Vscat -\Vmif\ will be approximately twice the evaporation velocity of the emitting gas. That  evaporation velocity will be \Vpdr -\Vmif. We adopt \Vpdr = 27.5$\pm$1.5 \kms (from \Cii ) and 
calculate expected \Vscat -\Vmif\ values of {\bf IF} 0 \kms, {\bf Near-IF} 10$\pm$2 \kms, {\bf Low-IZ} 15$\pm$2 \kms , {\bf High-IZ} 23$\pm$3 \kms.  The observed values of 13.2$\pm$0.6 \kms\ and 20.1$\pm$2.6 \kms\ for the two outer zones are in good agreement with the assumption of the \Vscat\ components being due to backscattering. A \Vscat\ component is not seen for \oi , which is not surprising because any backscattering component would be buried under the \Vmif\ component. We do not see a \Vscat\ component for \sii\ in the UVES spectra. However, there is one at \Vscat -\Vmif\ = 15.5$\pm$3.2 \kms\ for the higher signal to noise ratio HIRES sample, which is in marginal agreement with the expected value of 10$\pm$2 \kms , although there is a caveat to be stated above that the HIRES samples may include different tilts of the {\bf IF}. However, this caveat may not apply since \Vmif\ for \oi\ is very similar for HIRES (27.7$\pm$1.8) and UVES(28.1$\pm$0.1). 

In a fundamental paper on the expansion of a photoionized blister of gas, Henney et al. \citep{hen05} explained the blueshifts of lines arising from the MIF of the Huygens Region. The predictions of their model for a slightly concave PDR agree well with the velocity gradients seen in the strongest emission lines and in the various stages of ionization we study in this paper. 

The above conclusions are essentially the same as in earlier studies \citep{ode20a} that examined only \nii\ and \oiii\ spectra from a number of locations. What this study has done is to expand the list of emission lines, increasing the accuracy of testing  and confirming the photo-evaporation 
model of the MIF and backscattering from the PDR. 

\begin{figure}
 \includegraphics
[width=3.9in]
 {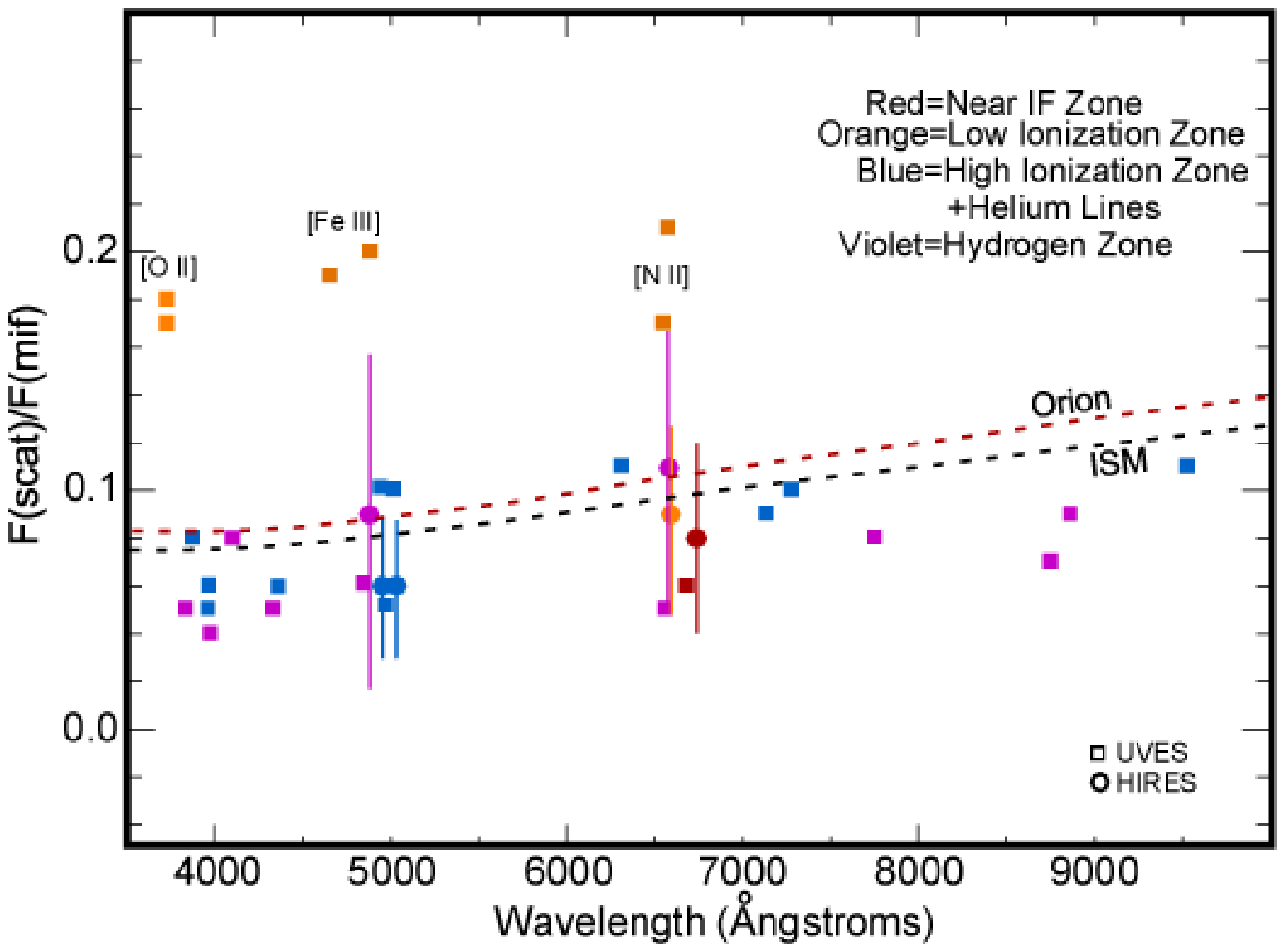}
\caption{The ratio of backscattered light to the source of that line's emission (\Fscat /\Fmif) is shown as a function of wavelength. Determination for the UVES \citep{md21} sample are depicted with small boxes and the HIRES samples \citep{hen99} as filled circles. Error bars for the HIRES values were determined from the scatter of values from typically 20 separate spectra.  The UVES error bars were determined using the IRAF task ``splot''. When no error bar is shown, it is no greater than the size of the symbol. The dashed lines indicate the expected relation for backscattering from an optically thick PDR (Section~\ref{sec:possible}).}
\label{fig:ratios}
\end{figure}

\subsection{Characteristics of  the \Fscat /\Fmif\ ratio}
\label{sec:ratio}

The \Fscat\ component of a spectrum is formed by backscattering in the optically thick PDR. This means that the \Fscat /\Fmif\ ratio is determined
by the albedo and backscattering phase-function of the dust particles in the PDR, and possibly the distance between the emitting and scattering layers.
Previous studies of the red shoulder of \nii\ and \oiii\ emission lines found a wide range of \Fscat /\Fmif\ values with a range of near zero to about 0.2.
%This ratio must be determined by both the separation of the emitting and scattering layers and the backscattering albedo of the PDR particles.
One of the goals of this study was to characterize this ratio over a wide range of wavelengths and ionization states.

In Figure~\ref{fig:ratios} we present the results for the UVES and HIRES spectra, where it is important to recall that the UVES values are from a single sample and the HIRES values are the average of about twenty spectra within five samples.
We detected a \Vscat\ component in all of our zones except {\bf IF} and {\bf Near-IF}. There the \Vmif\ values are so close to \Vpdr\ that any expected backscattering would be hidden 
under the red shoulder of the \Vmif\ profile.
The data are coded by symbols (filled boxes for UVES, filled circles for HIRES) and by colors for the emission zone. The error bars for the HIRES data are much larger, which probably reflects the variety of conditions in the multiple samples in five areas; however, their average values are similar to multi-wavelength values for the UVES sample.

The striking outlying values all arise from the UVES {\bf Low-IZ}. Within this set the much weaker \feiii\ lines agree with the \nii\ and \oii\ lines. It is possible that the small velocity separation of the \Vmif\ and \Vscat\ components prohibits accurate measurement of the \Vscat\ component in the red shoulder of the \Vmif\ component. However, the strength of the \Vscat\ components shown in Figure~\ref{fig:profiles} and the small uncertainties of the derived deconvolution argue that the ratios are accurate. It may be that the small separation of the {\bf Low-IZ} and the PDR is the cause, although to the first order this should not be a factor in emission and scattering from two plane parallel layers. Certainly the {\bf Low-IZ} zone is much closer to the PDR than the {\bf High-IZ}+{\bf He-Lines} and the {\bf H-Z}. The layer emitting \nii\ has a characteristic thickness of 0.003 pc (corresponding to 1\arcsec\ in the plane of the sky) and the \oiii\ characteristic thickness is about 0.06 pc (corresponding to 20\arcsec), as derived in \citet{ode01}. In addition, the {\bf Low-IZ} zone is highly structured, as shown by the much greater detail seen in images made in the low ionization states, whereas the higher ionization layers are smoother. 

We can only conclude that the much higher values of \Fscat /\Fmif\ are likely to be real, even though the reasons are not established. 
%Since the \oii\ and \feiii\ emission arise from the same narrow layer as the \nii\ emission, it is not surprising that all three ions show the same values of \Fscat /\Fmif .

\subsubsection{A possible wavelength dependence of the backscattering}
\label{sec:possible}

Any wavelength dependence of the \Fscat /\Fmif\ ratio may be a useful diagnostic of the nature of the particles in the PDR. The line of sight reddening to field stars is determined by a combination of absorption and diffuse scattering of interstellar particles, with a well established result that the extinction curve in the outer layers of the Orion Nebula being flatter than in the general interstellar medium. The easiest and most common interpretation is that foreground Orion particles are larger
than those usually encountered {\citep{bal91}. In the case of our observations the \Fscat /\Fmif\ ratio becomes a measure of backscattering by particles. 

In Figure~\ref{fig:ratios} one sees that for both the {\bf Near-IF} and the other zones that there are systematic patterns. For the {\bf Near-IF} points the value of \Fscat /\Fmif\ is essentially constant, with a hint of a drop at the 3727 \AA\ \oii\ doublet. For emission lines from the other zones, we see that \Fscat /\Fmif\ is again nearly flat with a believable indication of a drop in the ratio at the shortest wavelengths.  

\subsubsection{Expected wavelength dependence}
\label{sec:expected}

Studies of scattered starlight have been concentrated on the Merope reflection nebula  in the Pleiades star cluster \citep{ode65b,gibsonII}, where it was found that the scattered continuum 
was much redder than the illuminating star. However, this object is optically thin in visual wavelengths and the scattering layer is in the foreground of the star \citep{gibsonII}. This invalidates any 
comparison with Orion because dust in Orion's PDR must be optically thick and the scattering layer lies beyond the source of the emission.  

\citet{hen98} initially modeled scattering of light in Orion's PDR, extending those calculations in Appendix B of a later paper \citep{fer12}. There he established that the principal dependency was the single-scattering albedo and the scattering assymetry.  In a private communication he reports calculating the expected wavelength dependence for particles typical of the general interstellar medium (reddening ratio R$\rm_{V}$ = 3) and the foreground material reddening the starlight and nebular emission (R$\rm_{V}$ = 5). With his permission we have added these predictions to Figure~\ref{fig:ratios}. There is a remarkable general agreement of his predictions with the observations (except for the  {\bf Low-IZ} Zone, where those ratios are anomalous (Section~\ref{sec:ratio}).

\begin{deluxetable*}{lccccccc}
\floattable
\setlength{\tabcolsep}{0.02in}
\tabletypesize{\scriptsize}
\tablecaption{Summary of ELBC values, including earlier studies 
\label{tab:ELBCsummary}}
\tablewidth{0pt}
\tablehead{
\colhead{Line}&
%\colhead{$\rm \lambda _{o}$ (\AA)} &
\colhead{Earlier Source} &
\colhead{Spectograph} &
\colhead{Resolution}    &
\colhead{ELBC(Earlier*)} &
\colhead{ELBC(UVES**)} &
\colhead{ELBC(HIRES***)} &
\colhead{ELBC(ave$\dagger$)}}
\startdata
%I accidentally erased the FEED and HIRES entries and can't find a backup.
%\sii, 6731                        &\citet{ode03}              & HIRES                     & 6.2 \kms           & 11.3$\pm$2.4\\
\oi , 6300 \AA\                   &\citet{ode92a}              &FEED                    &4.7 \kms            & 8.9 (1)                     &7.9$\pm$0.2(4)       &10.3$\pm$1.8(3)    &8.9$\pm$1.2 \\
%~~~~~~~"                         &\citet{wen93}               & HIRES                 & 6.2 \kms          & 9.0$\pm$2.1\\
%~~~~~~~"                         &\citet{gar08}               &SPM***                   & 10 \kms           & 11.5$\pm$2.6(2)      &---                        &---                               & \\
\sii, 6731 \AA\                  &---                                &---                                &---                     &---                            &18.8$\pm$0.1(4)   &13.6$\pm$2.5(2)   &17.1$\pm$2.7\\
%\nii, 658.3 \AA\                  &\citet{ode03}              & HIRES                  &6.2 \kms           & 10.6$\pm$1.4\\
\nii, 6583 \AA\                        &---                          & ---                             &---                     &---                         &13.2$\pm$0.1(4)    &11.1$\pm$1.6(2)   &12.5$\pm$1.1\\
\oii, 3729 \AA\                    &\citet{jones92}            & FEED                   & 4.7 \kms        & 10.5$\pm$2.5(2)      &12.6$\pm$0.0(4)        &---                          &11.9$\pm$1.1\\
%\Ha , 656.3 \AA\                &\citet{ode03}              &HIRES                  & 6.2 \kms           & 19.5$\pm$1.3\\
%\Siii , 631.2 \AA\                &\citet{ode03}              & HIRES                  & 6.2 \kms          & 11.8$\pm$1.9\\
\Siii, 6312 \AA\                  &\citet{wen93}             &FEED                  & 4.7 \kms      & 11.0(1)          &10.3$\pm$0.2(4)    &---                            &10.4$\pm$0.3      \\
\oiii, 5007 \AA\                  &\citet{hoc88}               &FEED                    & 4.0 \kms          & 8.6$\pm$0.1(4)        &6.7$\pm$0.1(4)       &11.2$\pm$2.7(2)     &8.4$\pm$1.7\\
%~~~~~~~"                         &\citet{ode03}               & HIRES                 &6.2 \kms            & 13.0$\pm$3.7\\
%~~~~~~~"                         &~                                 &~                            & ~                     & ~                           &---                         &---                               &\\
%\Ha , 6563 \AA\                       &---                        &---                            &---                      & ---                  &11.0$\pm$0.1(4)     &19.7$\pm$3.4(1)   &12.7$\pm$3.9\\
%\oiii, 495.9 \AA\                   &\citet{ode03}               & HIRES                 & 6.2 \kms           & 11.3$\pm$2.1\\
%\Hb, 486.1 \AA\                   &\citet{ode03}               & HIRES                  & 6.2 \kms          & 18.8$\pm$2.2\\
 \enddata
%round off ratios to hundredths, do better ids on H
~$\dagger$ Weights are given in parentheses.

*The Coud\'e Feed Spectrograph at the KPNO observatory \citep{hoc88}.

** The UVES Spectrograph at the Very Large Telescope.

***The HIRES Spectrograph on the Keck-I 10-m telescope.

%**The Echelle Spectrographs on the KPNO and CTIO 4-m telescopes.

%***The MES-SPM Spectrograph on the Mexican National Observatories San Pedro Mart\'ir 2.1-m telescope.

\end{deluxetable*}

\section{The extra line-broadening component}
\label{sec:elbc}

\subsection{Earlier discussions of the ELBC}
\label{sec:earlier}

There have been multiple earlier studies with modern CCD detectors that have characterized the FWHM in various emission lines and over a range of spectrograph resolutions.
Some results come from studies pursuing interpretation of fine scale velocities of the MIF in hopes of explaining these as evidence of turbulent motion (\citep{hoc88},\citep{jones92},
\citep{ode92a},\citep{wen93}.

These results have been summarized in an earlier discussion of the ELBC \citep{ode17a}, but a re-evaluation is presented here because we now limit our HIRES data to the inner parts of the Huygens Region and have \Wobs\ values for many additional ions and lines. We gather in Table~\ref{tab:ELBCsummary} the data obtained in the same region as our
UVES and HIRES samples. This table also presents the results from the current study and a weighted average of ELBC from all the studies for each line. We do not use the \Wobs\ values for several lines from \citet{gar08} because they adopt a fundamentally different method of analyzing the line profiles.

\begin{figure}
 \includegraphics
%[width=\columnwidth]
[width=3.4 in]	
 {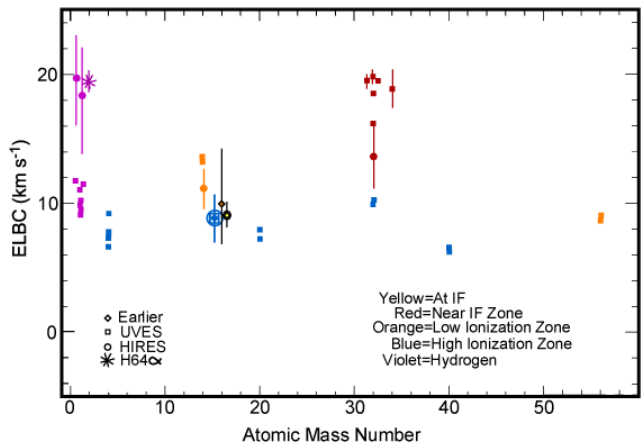}
\caption{The unexplained ELBC values determined by quadratic extraction from the observed \Wobs\ as explained in Section~\ref{sec:elbc} are presented. In order to identify any dependence of ELBC on the zone giving rise to the line, the data have been color coded, with the symbols having the same meaning as in Figure~\ref{fig:ratios}. Entries for oxygen from the same ionization zone were averaged.  Where the oxygen results were crowded, the plotted A values were shifted by 0.5. The error bars were determined as in Figure \ref{fig:ratios}. The data point for hydrogen at 19.6$\pm$0.9 \kms\ is from a study of H64$\alpha$ by \citet{wilson97}, with ELBC calculated in \citet{ode03}. }
\label{fig:widths}
\end{figure}

\subsection{Derivation of the ELBC}
\label{sec:derivation}

As noted in Section~\ref{sec:broadening} the observed line width (as measured by its FWHM) depends on several factors: the Atomic Mass (A), the electron temperature, any fine-structure component (\Wfs), the velocity resolution of the spectrograph
employed (R). The earliest high-resolution optical studies of the strongest emission lines indicated that these factors failed to explain the observed FWHM (\Wobs), which lead to the introduction of an additional broadening term (the ELBC). Equation~\ref{eq:Wobs} can then be rewritten to isolate the ELBC component as
\begin{equation}
\label{eq:elbc}
\rm ELBC =\left[W_{obs}^{2} -438.1/A -R^{2}-W_{fs}^{2}] \right]^ \frac{1}{2}
 %\rm W_{obs} = \left[424.1/A +ELBC^{2} + R^{2} +W_{fs}^{2} \right]^ \frac{1}{2}
\end{equation}
where an electron temperature of 9200 K \citep{md21} is assumed. We adopted the fine-structure line-broadening constants of \citet{gar08} for \Ha , \Hb\ and \Hgamma.
We show the derived ELBC values in Table~\ref{tab:AllHIRESLines} for the HIRES spectra, Table~\ref{tab:ELBCsummary} for various previous high-resolution optical studies, and Table~\ref{tab:AllUVESzones} for the UVES Zones.  We also show the derived ELBC values from the various emission line studies and an H64$\alpha$ study \citep{wilson97,ode03} in Figure~\ref{fig:widths} .

\subsection{Derived ELBC values}
\label{sec:derived}

We have used the procedure described in the preceding section to derive ELBC values from the \Wobs\ determination for all all the samples under consideration. Although most fall into a value near 9 \kms , there are some anomalies that must require a different interpretation.

\subsubsection{ELBC in the Hydrogen Lines}
\label{sec:hydrogen}

The most difficult ion for extracting ELBC is hydrogen (A = 1) because of its large thermal broadening and the presence of a fine-structure component. The large thermal component means that the derived ELBC is sensitive to the electron temperature. Therefore the clearest values of ELBC for hydrogen lines is for the UVES data where the same data set was used for deriving the electron temperature, so any uncertainty of the thermal component should be small.

The average value of ELBC for the UVES hydrogen lines (10.5$\pm$0.9 \kms) falls near the values derived from other ions (about 9 \kms).

\subsubsection{Other Ions}
\label{sec:others}

At helium (A=4) we have only the results from UVES. Their ELBC values are grouped around 7.8$\pm$1.1 \kms. The helium results are expected to be accurate as the thermal broadening component is smaller and the \heI\ lines arise only from the {\bf High-IZ} 

The oxygen values cluster within 1 \kms\ around 9.5 \kms. Oxygen is present in all of the ionization zones (\oi\ at the {\bf IF}, \oii\ in the {\bf Low-IZ}, and \oiii\ in the {\bf High-IZ}.

%We have not plotted the \clii\ (A=34) value of 18.7$\pm$0.7 \kms\ because of the complex origin of the emission line at 9124 \AA\ as discussed in Section~\ref{sec:velocities}. 

The results for ELBC from \neoniii\ (A=20), \ariii\ (A=40), and \feiii\ (A=56) all lie near the values for  UVES (\hi), \heI , and oxygen. The \nii\ (A=14) values from both HIRES and UVES lie slightly higher than this grouping. Because the line (6583 \AA) is well resolved from the nearby \Ha\ line (6563 \AA), contamination should not be a problem.

Although ELBCs for the \siii\ lines (6313 \AA\ and 9531 \AA ) lie near the values from most of the other ions and lines, there is an unexpected wide range for the \sii\ lines. These are discussed in Section~\ref{sec:anomalous}.

\subsubsection{Anomalous values of \hi , \sii, and \clii\ lines}
\label{sec:anomalous}

The ELBC results for \hi\ in the radio H64$\alpha$ \citep{wilson97} and HIRES samples are quite different from those in the UVES sample. The UVES region is quiescent and unaffected by the high-ionization shocks in HH~529 that lie well into the foreground, whereas the HIRES samples are all from samples with proplyds or high-velocity features.

One thing that is different from the other observations is the spatial resolution. \citet{wilson97} employed a beam width of 42\arcsec, and the HIRES slit samples were about 7\arcsec\ long, while our UVES sample was 
5\farcs2 long. Since the HIRES and UVES samples are of similar sizes, the explanation probably does not lie with the spatial resolution but in the nature of the region in our UVES sample.

The HIRES and H64$\alpha$ values would come into agreement with the UVES data if the electron temperature was about 13000 K, which is much higher than theoretically expected. This is because at a higher temperature the thermal component correction of the broadening would be larger and for a fixed observed W, the derived ELBC would be reduced. Although the hydrogen emission comes from throughout the ionized gas, because it is seen in recombination, the emissivity is weighted towards lower temperatures. Therefore it is unlikely that the large ELBC values are due to higher electron temperatures.

The HIRES result for 
6731 \AA\ (13.6$\pm$2.5 \kms ), could be explained by higher electron temperatures in those samples. However, this cannot be invoked for explaining the high UVES 6731 \AA\ values, because the temperature there is accurately determined. The same can be said for the \clii\ ion (18.7$\pm$0.7 \kms), as determined from the 9124 \AA\ line. The fact that both \sii\ and \clii\ emission arise from the {\bf Near-IF} argues for a common explanation. This could
be much higher turbulence in this transition zone or more intense magnetic fields where the condition of a frozen-in field is established.

William J. Henney in a private communication has pointed out that the UVES sample lies in a quiescent region, as noted in Section~\ref{sec:existing}. In contrast, all of the HIRES samples lie 
in regions that systematically show higher values of the ELBC (c. f. Figure~16 of \citet{gar08}). This could account for the high value of ELBC found from the HIRES \Ha\ data and the
H64$\alpha$\ results at a resolution of 42\arcsec . This could be also be true for the somewhat high values in the HIRES \nii\ and \sii\ data. However, this cannot explain the anomalously high values from the UVES \sii\ and \clii .The variation between sample regions emphasize that the UVES values are to be employed for an explanation of the ELBC.

\subsection{Explanation of the ELBC}
\label{sec:explanation}

The problems that confront us are to explain the cause of the ELBC that we find in most samples (the band of values at about 9 \kms\ across Figure~\ref{fig:widths}) and the high values of ELBC 
that are seen in \hi\ in the HIRES samples, in the H65$\alpha$ sample, and in the \sii\ and \clii\ UVES samples.

\subsubsection{Earlier attempts to explain the ELBC}
\label{sec:earlier}

\citet{hen05} investigated the role that the velocity gradient in a photo-evaporation flow has in broadening an emission line and producing the ELBC. They conclude that photo-evaporation contributes little to the line-broadening and then evaluated the role that might be played by \alf\ waves. They concluded that these waves could explain their value for ELBC in \siii\ (9 \kms) if
the magnetic field was approximately 10$\rm ^{-3}$ gauss.  but concluded that in general \alf\ waves do not provide a natural explanation for the line-broadening.

%that their models agree with their ELBC for \siii\ (that found 9 \kms\ for the spectra later published in \citet{gar08}, we find 10 \kms), but are in strong disagreement for \oi , where they found 12 \kms\ (we find 8 \kms). Their models predict essentially a zero value for the \oi\ ELBC since that line arises at the ionization front. They also examine the possibility that the broadening is caused by Alfv\'en waves, but consider these unlikely to produce the extra line width.}

The ELBC was also explored in detail in Section 6.1.2 of \citet{ode17a}, where they evaluated the role of the expansion velocity of \nii , considering both quadratic and linear addition. They too concluded that evaporation flow was not a major contributor to the ELBC. Their data set used HIRES results and the \citet{gar08} atlas spectra, exploring grouping of ELBC according to ionization zones and forbidden-versus recombination lines. The homogeneity and multiplicity of lines in the current study render that discussion moot, except for the qualitative argument that the answer 
could be due to photo-evaporative flow off of dense knots in the PDR, as modeled by \citet{mel06}, although this was not considered quantitatively in any detail. 

\subsubsection{The ELBC may be explained by \alf\ waves}
\label{sec:favored1}
Magnetic fields in \hii\ regions and the associated PDR's have previously been established, a good review appearing in \citet{fer09}. \citet{jess} successfully explained the non-thermal line widths in the lower layers of the Sun's atmosphere as being due to \alf\ waves and \citet{roshi} determined in 14 samples of \hii\ regions that \alf\ waves explained the non-thermal components of singly ionized carbon radio recombination lines. 

It is well established that there are strong magnetic fields in the Huygens Region. In their examination of absorption lines in the closer (to \tC) Veil component (called layer B), \citet{tom89} and \citet{tom16} found 
values for the line-of-sight magnetic field about 50 $\rm \mu$G, far above the usual values for the interstellar medium. In an analysis of the energy parameters, they found that most of the energy was in the magnetic field. This is similar to the conclusion of \citet{abel16} that layer B in the Veil is in rough equipartition.

The presence of a strong magnetic field in the PDR, which is much closer to the MIF than the Veil, has been derived from observations of polarization in far IR thermal emission from aligned dust grains \citet{chuss}. Their ``\hii\ region'' sample, which is close to our UVES sample, found a plane-of-the-sky value of 0.29$\pm$0.03 mG. This value is probably too low, as they assumed that the thermal grain emission comes from a layer 0.14 pc thick. \citet{goi16} show in their ALMA study of HCO$\rm^{+}$ in the highly tilted Bright Bar that the apparent thickness of the PDR is 30\arcsec , which corresponds to a distance of 0.06 pc at an adopted distance of 388 pc \citep{mk17}. Under the assumption that the PDR thickness is the same in both the Bright Bar and their ``\hii\ region''sample, the true derived magnetic field in the plane of the sky is twice as large as their value and the 3-D field will be even larger.

\citet{hen05} give the formula for \alf\ wave velocities \Va$^{2}$ =B$^{2}$/4$\pi \rho$, where $\rho$ is the mass density of the gas. Assuming that the gas is fully ionized and dominated by the mass of hydrogen, this equation becomes \Va$^{2}$ (\kms) = 4.79$\times$10$\rm^{12}$  B$^{2}$/\Ne, where \Ne\ is the electron density. For \Ne\  about 10$\rm ^{4}$ electrons/\cmq, this becomes \Va$^{2}$ (\kms) = 4.79$\times$10$\rm^{8}$ B$^{2}$. If B = 1 mG, then \Va $\approxeq$ 21.9 \kms. This is in approximate agreement with ELBC = 9 \kms, similar to the conclusion of \citet{hen05} for his \siii\ line. However, this approximate agreement does not make it clear that it can explain the constancy of ELBC across the ionization zones since we have no a priori knowledge of the variations in B with ionization.  

\citet{tom16} presents equations for the total thermal and magnetic field energy densities as E$_{therm}$ = (3/2) nkT and E$_{mag}$=B$^{2}$/8$\pi$. The ratio of the magnetic and thermal 
energy densities ({\bf $\beta$}) in a fully ionized volume will then be {\bf $\beta$} = B$^{2}$/(12$\pi$kT\Ne). Substituting the ratio B$^{2}$/\Ne\ into the \Va\ equation gives \Va$^{2}$ = 3 k T {\bf $\beta$}/m$_{H}$. Inserting numerical values for the constants and assuming that the electron temperature is 10$\rm ^{4}$ K, then expressing the velocity in the convenient units of \kms\ gives \Va$^{2}$ =249$\times$ {\bf $\beta$}.
Since the ELBC component is 9 \kms , this argues that {\bf $\beta$} = 0.33, and the magnetic and thermal energies are about equal.
If the ELBC is caused by \alf\ waves, and ELBC is nearly constant in all the ionization zones, this argues that {\bf $\beta$} is about equal throughout. This means that the ratio B$^{2}$/\Ne\  is constant as \Ne\ drops with increasing distance from the PDR. How the observed strong magnetic field in the PDR is coupled to the inferred magnetic field in the outflowing ionized gas and the inferred B$^{2}$/\Ne\ constancy is not addressed here.

In the models of Orion created by \citet{hen05} the effects of magnetic fields in an expanding photo-ionized layer were considered and in some regions were quite important. A much more elaborate set of \hii\ region models, now giving full consideration of magnetic fields, is described in \citet{art11}. Although the latter study employed the most complete considerations of the MHD processes invoked by the presence of a magnetic field, none of those models simulates the Huygens Region specifically. A magnetic-field-free statistical analysis of ELBC and radial velocities by \citet{art16} demonstrated that the observed ELBC was about 1.5 times larger than expected from the point-to-point velocity variations, but that this could be explained by turbulence in the ionized regions. Since turbulence must be present, then the derived \alf\ wave velocities are upper limits.

We consider the explanation of the ELBC to be an open subject. \alf\ waves are attractive because of the simplicity of a constant near energy equilibrium of the thermal gas and the magnetic field. Its relation to the strong field in the PDR may be due to its being frozen-in with the plasma. In this case the magnetic field lies are trapped into the expanding ionized gas. %Although Alfv\'en waves have been invoked to explain time dependent variation in the width of \Ha\ lines in the solar chromosphere\citep{jess} but 
The exact nature of Alfv\'en waves in conditions within the interstellar medium remain a long-standing problem in the discipline.
 
\subsubsection{Explanation of the anomalously high ELBC values}
\label{sec:favored2}

In Section~\ref{sec:anomalous} we present the arguments for the extraordinarily large ELBC values for the HIRES and H64$\alpha$ being due to those samples coming from regions of larger ELBC values, whereas the lower values for the UVES sample are because that region is genuinely more quiescent.  Within a model invoking the importance of \alf\ waves, the extraordinarily large UVES ELBC values for \sii\ and \cliii\ could be due to dominance by the magnetic field in the narrow region of ionized gas closest to the PDR.

    \section{Conclusions}
    \label{sec:conclusions}
  $\bullet$ The observed velocity gradient agrees with the expectation of viewing a nearly flat-on, photo-evaporating ionization front.
  
  $\bullet$ The separation of the \Vmif\ and \Vscat\ is consistent with the red shoulder of the emission lines being due to backscattering in the PDR.
  
  $\bullet$ The backscattering component is stronger when the emitting layer is close to the PDR, being about one-half as strong when the emitting layer is further away in the \ozone\ Zone.
  
  $\bullet$ There is a suggestion of a wavelength dependence of the strength of the backscattered light, it being somewhat lower at shorter wavelengths. 
  
  $\bullet$ The slight wavelength dependence of the backscattering is consistent with models adopting characteristics of either the interstellar medium or the Orion cluster.
  % very different from that in the Merope Nebula, which is formed when the cool stars of the Pleiades Galactic Cluster is passing through an interstellar cloud having no relation to the creation of the cluster members.
  
  $\bullet$ An extra component contributing to the emission line widths is certainly present, generally being about 9 \kms .
  
  $\bullet$\ Although this extra line-broadening of emission lines  can be attributed to turbulence, it can also be explained by \alf\ waves in a medium where the ratio of magnetic energy and thermal energy is constant.
  
 % $\bullet$ The constancy of the extra broadening component across most of the MIF argues that the magnetic field over thermal gas energies is nearly constant. 
  
  $\bullet$  Anomalously high line-broadening is encountered in the {\bf Near-IF} zone in the UVES sample and in the HIRES hydrogen and radio H64$\alpha$ samples, the former are probably due to dominance by the magnetic field and the latter due to the HIRES and H64$\alpha$ sampling more turbulent regions.

 \section*{acknowledgements}
 
We are grateful to W. J. Henney of the Instituto de Radioastronom\'ia y Astrof\'isica, Universidad Nacional Aut\'onomia de M\'exico, Campus Morelia, M\'exico for providing the theoretical models incorporated in Figure~\ref{fig:ratios} and for his illuminating remarks on an earlier draft of this paper.
 Thanks are also due to David T. Chuss of Villanova University for useful discussions. JEM-D thanks the Instituto de Astrof\'isica de Canarias for support under the Astrophysicist Resident Program and acknowledges support from the Mexican CONACyT (grant CVU 602402). 
 
The HIRES data presented herein were obtained at the W. M. Keck Observatory, which is operated as a scientific partnership among the California Institute of Technology, the University of California and the National Aeronautics and Space Administration. The Keck Observatory was made possible by the generous financial support of the W. M. Keck Foundation. 

 \section*{Data Availability}

The UVES data were from observations collected at the European Southern Observatory, Chile, proposal number ESO 092C-0323(A), on the nights of November 28 and 29, 2013 using the Very large Telescope in Cerro Paranal, Chile and are available in their archives. The data are described in Appendix D2 of \citet{md21} and their processing is described in Section 2 of the same publication. The Keck HIRES observations and processing are described in \citet{hen99} and were obtained on the nights of  December 5 and 6, 1997 and are available through the Keck Observatory Archives.

%\newpage                                           

\end{document}